Effect of surface hydrogen on the anomalous surface segregation behavior of Cr in Fe rich Fe-Cr alloys

Michèle Gupta[1, $], Raju P. Gupta[1], and Maximilien Levesque[2]

[1]Laboratoire de Thermodynamique et Physico-Chimie d'Hydrures et Oxydes, Université de Paris-Sud, 91405 Orsay, France

[2]École Normale Supérieure, Département de Chimie, UMR 8640 CNRS-ENS-UPMC, 24 Rue Lhomond, 75005 Paris, France, and CNRS, UPMC Université Paris 06, ESPCI, UMR 7195 PECSA, 75005 Paris, France

Abstract

The segregation behavior of Cr in dilute Fe-Cr alloys is known to be anomalous since the main barrier for surface segregation of Cr in these alloys arises, not from the topmost surface layer but, from the subsurface layer where the solution energy of Cr is much more endothermic as compared to the topmost surface layer. The Fe-Cr alloys are candidate structural materials for the new generation of nuclear reactors. The surfaces of these alloys will be exposed to hydrogen or its isotopes in these reactors, and although hydrogen is soluble neither in Fe nor in Fe-Cr alloys, it is known that the adsorption energy of hydrogen on the surface of iron is not only exothermic but relatively large. This clearly raises the question of the effect of the hydrogen adsorbed on the surface of iron on the segregation behavior of chromium towards the surface of iron. In this paper we show, for the first time, on the basis of our ab initio density functional theory (DFT) calculations, that the presence of hydrogen on the surface of iron leads to a considerably reduced barrier for Cr segregation to both the topmost surface layer and the subsurface layer but the subsurface layer still controls the barrier for surface segregation. This reduction in the barrier for surface segregation is due to the nature of the Cr-H couple that acts in a complex and synergistic manner. The presence of Cr enhances the exothermic nature of hydrogen adsorption that in turn leads to a reduced barrier for surface segregation. These results should be included in the multiscale modeling of Fe-Cr alloys.

PACS Numbers: 61.82.Bg, 64.75.Nx, 75.50.Bb, 68.43.Bc

# I. INTRODUCTION

The Fe-Cr alloys with low concentration of Cr, ~10 atomic percent, are being considered as structural materials for the next generation of fusion and fission reactors [1-4] due to their better mechanical properties, greater resistance to radiation damage and helium-induced swelling, and corrosion resistance, among others. A considerable amount of work [8-14] is being done on the various aspects of these alloys, both experimental and theoretical multiscale modeling, in order to gain insight into their behavior from a fundamental point of view. The segregation of Cr to the surface of these alloys plays a major role in controlling many key properties, including corrosion resistance. Although, hydrogen is an extremely important element, much less attention has been paid to its effects on the properties of these alloys, in comparison to the case of helium where the literature is abundant concerning its detrimental effects. In fusion reactors, these alloys will be directly exposed to (H,T) plasma. The hydrogen can also be released under accidental conditions exposing the surface of these alloys to hydrogen. It also plays a key role in controlling the corrosion resistance and the mechanical properties of materials. The effect of the presence of hydrogen on the surface of these alloys on the segregation behavior of Cr has not been investigated.

In solid solution forming alloys, such as Fe-Cr, the energy of solution of the segregating element on the surface is less exothermic than in the bulk, and hence energy is required for the element to segregate to the surface. In most binary alloys, the energy of solution becomes more exothermic gradually from the surface towards the subsurface layers until it reaches its bulk value far from the surface. Thus the segregation energy, the maximum energy required for segregation, can be determined in a straightforward manner from the difference in the solution energies at the topmost surface layer and in the bulk. This nice picture of surface segregation, unfortunately, breaks down for dilute Fe-Cr alloys since the energy of solution of Cr in the subsurface layer, just below the topmost surface layer, is not only not gradual nor exothermic but is rather large and endothermic [10-14]. This means that the maximum energy for the segregation of Cr, the segregation energy, is not determined by the topmost surface layer but rather by the subsurface layer, and this has to be necessarily included in all thermodynamic modeling involving Cr segregation. This anomalous segregation behavior of Cr has been attributed [14] to a complex interaction that involves the antiferromagnetism of Cr and the ferromagnetism of Fe that modify considerably the nature of atomic relaxations, resulting in the modification of the electronic structure in such a manner that it does not favor the incorporation of Cr in the subsurface layer.

Surface segregation is a complex phenomenon [15-19] in concentrated alloys, and unexpected behavior can be observed in certain systems, for example NiPt alloys [16]. Further, in many magnetic materials, magnetism is known to alter dramatically the physical properties, leading to, in many cases, industrially important applications such as the INVAR effect in Fe-Ni alloys [17], and the giant magnetoresistance in Fe-Cr multilayers [18]. Thus it is not unexpected that the magnetic interactions between Fe and Cr atoms can result in an unexpected segregation behavior. The presence of hydrogen on the surface of Fe-Cr alloys is expected to further modify the segregation behavior of Cr through its coupling to Fe and Cr atoms. We have therefore investigated, using ab initio electronic structure calculations, the effect of hydrogen, adsorbed on the topmost surface layer, on the segregation energy of Cr. As in our previous work [14], we will not be dealing with dilute Fe-Cr alloys but we choose instead the case of Cr as an impurity in Fe, which is relevant to Cr segregation in dilute Fe-Cr alloys. We have selected for this purpose the (1 0 0) surface of ferromagnetic Fe which is known to be the most stable surface layer. The computational details are given in Section II. The results on the adsorption energies of H with and without Cr are presented in Section III where the solution and segregation energies of Cr are also calculated. Concluding remarks are given in Section IV.

## II. COMPUTATIONAL DETAILS

The electronic structure calculations presented in this work were performed within the density functional theory (DFT) using the spin polarized version [19-21] of the Vienna Ab Initio Simulation Package (VASP) in the Generalized Gradient Approximation (GGA). The Projected Augmented Wave (PAW) potentials [22] were employed in conjunction with the PW91 version of GGA [23] to account for exchange and correlation corrections. The ferromagnetism of Fe plays a critical role in determining the properties of iron, and its inclusion is absolutely essential for a reliable determination of the defect properties in iron. The 16 outer electrons of Fe (8 semi-core electrons and 8 outer valence electrons), and 12 outer electrons of Cr (6 semi-core electrons and 6 outer valence electrons), were treated as valence electrons in the present computations. The semi-core states were not included in previous calculations, but can play an important role in the determination of certain properties, such as phonon spectra which are sometimes found to be imaginary without the inclusion of these semi-core electrons in an otherwise perfectly stable lattice. We consider in this work the Fe (1 0 0) surface since this surface is found to be the most stable. We used a repeated slab geometry with dimensions ($2a$ $2a$ $4a$) along the a-, b-, and c-axes, separated by a vacuum layer of $4a$ (~11.5 Å) along the c-axis, where $a$ is the lattice constant of bcc Fe. The supercell thus contained 36 atoms (9 layers in the c-direction). An energy cutoff of 500eV was employed, and the results were found to be fully converged using a (7 7 1) k-point grid in the Monkhorst-Pack [24] scheme. We used the theoretically determined lattice constant ($a$=2.826 Å) of ferromagnetic Fe. This corresponds to treating both Cr and H in the infinitely dilute limit. The mirror symmetry (with respect to the central layer) along the c-axis was exploited when the H and Cr atoms were put on the surface. In other words, H or Cr atoms were symmetrically placed on each of the two surfaces. None of the layers was kept fixed, and the atomic coordinates were fully relaxed until the total energies were converged to $10^{-4}$eV or better, and the forces on the atoms were converged to better than $10^{-3}$eV/Å. The magnetic moment in bulk Fe was found to be 2.20 $\mu_B$, in good agreement with experimental measurements and previous calculations [7-14]. For pure bulk Cr, we obtained an antiferromagnetic ground state in the bcc structure with a magnetic moment of 0.97 $\mu_B$ and a lattice constant of 2.866 Å, which is only marginally larger than the lattice constant of ferromagnetic Fe. In fact, it is well known that Cr shows a spin density wave (SDW) behavior, but this SDW is quenched in the presence of impurities, and a regular antiferromagnetic behavior is observed. In our calculations, the magnetic moment of Cr was found to be aligned antiferromagnetically to those of Fe atoms in all cases. We mention here that test calculations were also performed with a larger slab with dimensions ($2a$ $2a$ $5a$) using a larger energy cutoff of 850eV, and a denser (9 9 3) k-mesh, using a harder version of the PAW potential for hydrogen, in order to be sure that the calculations are fully converged, and the results accurate. The difference in the energy of solution was less than 5meV compared to the results presented in this paper. The effect of the zero point energy (ZPE) contribution of hydrogen was neglected since the segregation energy in Eq; (5) given below, which is the main focus of this paper, depends upon the differences of the total energies in such a manner that if the ZPE contribution was the same in all cases, it will exactly cancel out in the segregation energy, and there will be no zero point energy contribution. Thus only the differences $\Delta$ZPE are relevant to the physics of the problem, and these are expected to be small. The exact treatment of ZPE requires extensive computations which are not warranted knowing that it will hardly affect the outcome of the results.

### III. RESULTS AND DISCUSSION

#### A. Adsorption energy of hydrogen on the surface of pure iron

We first discuss the results on the energy of adsorption of hydrogen (solution energy of H) on the (1 0 0) surface of pure Fe. The three possible interstitial sites for the adsorption of hydrogen on the (1 0 0) surface of Fe are shown in Fig. 1 where the corresponding bulk interstitial sites are also shown. In the bulk Fe, there are only two types of interstitial sites, octahedral and tetrahedral, denoted by the symbols O and T. In order to make connection with the surface sites, we split the octahedral sites in the bulk into two groups, those which are at the center of the faces, denoted O1, and those which are at the center of the edges, denoted O2, as shown in Fig. 1. The O1 sites correspond to the hollow sites (denoted H) at the surface. A hydrogen atom at this site is at the center in a square planar configuration formed by four Fe atoms on the surface at a distance of $0.707a$, but it also has one Fe atom at a closer distance, $a/2$, just below the surface in the subsurface layer. The O2 atoms of the bulk when viewed at the surface are now more properly termed as bridge sites (denoted B) since they are at the center of a bridge formed by two Fe atoms. Each H atom at this site has two Fe atoms at a distance of $a/2$ at the surface and two Fe atoms further, in the subsurface layer at a distance of $0.707a$. All of the tetrahedral sites on the surface (denoted T) are the same and each H atom at these sites has three Fe atoms, two on the surface and one in subsurface layer, at a distance of $0.56a$, compared to four in the bulk, for example, for the tetrahedral site situated in the subsurface layer. The last possibility is a tetrahedral-type site (in the bulk it has a tetrahedral coordination but not on the surface) just above the surface where the H atom has only two Fe atoms located in the surface plane. Note that there are four hollow sites on a (1 0 0) surface, and this corresponds to one hollow site per Fe atom. In this paper, we will be mostly concerned with the adsorption of a single hydrogen atom on each surface. In this case, the hydrogen adsorption energies, $E^H_{ad}$, at different interstitial surface sites can be calculated using the expression

$$E^H_{ad}= (E_{Tot}(Fe_N H_2) - E_{Tot}(Fe_N) - E_{Tot}(H_2))/2, \quad (1)$$

where $E_{Tot}(Fe_N H_2)$, and $E_{Tot}(Fe_N)$ are the total energies of a slab containing N Fe atoms and one H atom symmetrically placed on each surface (thus two H atoms per slab), and a slab without H atoms respectively, and $E_{Tot}(H_2)$ is the total energy of a $H_2$ molecule calculated with the VASP code. Thus the energy for the dissociation of a $H_2$ molecule has to be furnished before H adsorption can occur.

The most preferred site for H adsorption is found to be the fourfold coordinated hollow site (denoted as H), with adsorption energy of -0.405eV/H atom. However, the H atom at this site is not located in the surface plane of Fe atoms, but is instead displaced upwards (vertically) by 0.38 Å. This upward displacement is fairly easy to understand. The H atom at the O1-type octahedral site in the bulk is squeezed between two Fe atoms at a very close distance of $a/2$. On the surface in the hollow site position, one of the Fe atoms is no longer present due to the formation of the surface. This releases the constraint, and results in a much larger Fe-H distance, ~1.76 Å with the Fe atom just below it. We also investigated a position of an H atom slightly displaced inwards inside the surface but this converged to the final position obtained above. The magnetic moments at the four Fe sites closest to H atom are lowered only slightly to a value 2.75 $\mu_B$ from their value, 2.83 $\mu_B$, on the surface of pure Fe. The displacements of the H atom in the surface plane in the (1 0 0) and (1 1 0) directions from the hollow site were also attempted. The vertical distance from the surface did not change appreciably but the H atom did find a stable position at ~0.16 Å away from the hollow site. This clearly indicates that a shallow minimum energy surface exists around the fourfold coordinated hollow site in which a H atom can move freely. This picture is confirmed by our calculations for a H atom placed on a tetrahedral site on the surface plane. The hydrogen atom did not stay at this tetrahedral site but instead moved to the hollow site, but slightly displaced, ~0.16 Å, away from the hollow site at the same vertical distance as obtained for the H at the hollow site.

The tetrahedral sites also exist just underneath the surface at a vertical downward distance of $0.25a$ just below the bridge site. The H atom placed at this site prefers to move upwards towards the surface so that it is just beneath the surface at a distance of only 0.26 Å where its solution energy is only slightly exothermic, ~-0.06eV/H. In fact, the same configuration is also obtained by placing the H atom exactly at the bridge site on the surface plane. In this case, the H atom moves down the surface by ~0.26 Å from its initial bridge site position.

A final possibility for placing the H atom at a tetrahedral-type site is to place it a distance of $0.25a$ just above the surface bridge site. Again, just as in the case of the hollow site, the H atom gets displaced upwards to a vertical distance of 1.10 Å from the surface, to optimize the Fe-H distance, where its adsorption energy is considerably exothermic, ~-0.38eV/H, but still slightly less exothermic than at the hollow site, -0.405eV/H. Our results are in general agreement with those obtained by Jiang and Carter [25] using the VASP code and PAW potentials, although these authors used a smaller energy cutoff, 350eV, and a five layer slab in which three bottom layers were kept fixed

The discussion above clearly shows that the most favorable position for H adsorption on the (1 0 0) surface of Fe is the hollow site. We will therefore consider in detail only the hollow site in the discussion that follows. With the hydrogen at this site, a slightly larger contraction of the first interlayer separation, and a slightly smaller expansion of the next interlayer separation were obtained as compared to the pure Fe-slab calculations, as shown in Table I, in general agreement with the trend expected on the surfaces of transition metals [26]. Similarly, the magnetic moments of the Fe atoms at the topmost surface layer were only slightly reduced relative to those of the H-free slab. The most affected Fe site is the one just underneath the H atom in the subsurface layer where we obtain a magnetic moment of 2.21 $\mu_B$ as shown in Table I, which is much smaller than the value 2.32 $\mu_B$ obtained at the other Fe sites in the subsurface layer. However, these reductions in the magnetic moments are not really very large. This result is consistent with previous work [27] concerning the changes in the magnetic moments on Fe sites upon hydrogenation in the intermetallic compound $YFe_2$, where also only a slight reduction in the magnetic moment at the Fe site was found due to Fe-H interaction.

The adsorption energy, ~-0.405eV/H, of hydrogen at the surface of Fe is exceedingly exothermic, especially when one considers the fact that H is not soluble in bulk Fe. This result does not change significantly when all four hollow sites are occupied by the H atoms in which case we obtain a value of -0.414eV/H atom, but shows the attractive nature of the H-H interaction on the surface of Fe. In order to verify the fact that H is indeed insoluble in bulk Fe, the solution energy of H in bulk Fe was calculated by placing the H atom at the central layer. It was found that H prefers the tetrahedral site over the octahedral one by ~0.10eV but the solution energy of H at the tetrahedral site is, nevertheless, endothermic, +0.15eV. This means that H is more stable at the surface by ~-0.56eV relative to the bulk. This can be understood from the difference in the coordination of the H atom in the bulk and the surface, and the corresponding H-Fe distances. In the bulk, an octahedral position is not favorable for an H atom due to a very short Fe-H distance, $a/2$ (1.41A), with the two closest Fe atoms. In the tetrahedral site, an H atom has four nearest neighbor Fe atoms but at a much larger Fe-H distance, ~1.58 Å. This is the reason why the H atom prefers a tetrahedral site over the octahedral site in the bulk since a larger Fe-H distance, in the tetrahedral coordination provides an H atom more room for relaxation. On the surface it has only one Fe atom at a distance of $a/2$, just below it, and thus it has a much greater freedom for relaxation. This extra degree of relaxation plays a crucial role in the adsorption of H on the surface of Fe. As was stated above, the H atom is not located at the surface plane of Fe atoms but displaced upwards by 0.38 Å. Our calculations show that this upward displacement plays a major role in facilitating the adsorption of H since if we take the final relaxed configuration with H atom that includes the relaxation of the Fe atoms induced by H but move the H atom to the surface, ignoring completely the upward vertical displacement of 0.38 Å, the H atom does not adsorb at the surface since the adsorption energy in this case is endothermic, 0.24eV/H atom due to the short H-Fe distance, 1.36 Å. Similarly, if we put an H atom at the Fe surface (relaxed without H atom), and do not allow any relaxations induced by the presence of H, the H atom is not adsorbed at the surface of Fe either, since it has an endothermic adsorption energy, 0.13eV/H atom. On the other hand, if we move the H atom to a vertical distance of 0.38 Å found above, above the Fe surface the adsorption energy is exothermic, -0.387eV/H atom, only slightly less exothermic, by -0.018eV/H atom, than the value -0.405eV/H atom that includes the relaxation of the Fe atoms due to the H atom at the surface. Thus the prominent role in the adsorption of H in this case is played by the direct interaction of the H atom with the closest Fe atom, situated in the subsurface layer, at a distance of 1.72 Å, and a much weaker interaction with the four Fe atoms in surface layer which are further away at a distance of 2.04 Å although each of these Fe atoms is laterally displaced by ~0.03 Å from its ideal position due to the presence of H. The gain in energy due to these displacements is small, ~0.018eV/H atom.

### B. Solution and segregation energies of Cr in the presence of H on the surface

Given the large exothermic heat of adsorption of H on the surface of Fe, it is natural to find how the segregation behavior of Cr will be modified by the presence of H on the surface of Fe. Another related question is whether the presence of Cr in different surface layers will also modify the heat of adsorption of H on the surface of Fe through its interaction with the H atom on the surface. These results are presented below.

As was stated above, we exploited the mirror symmetry, with respect to the central layer, along the c-axis in this work. We put a single Cr atom in a given layer; there are thus two Cr atoms per slab. As in the case of pure Fe, the heat of adsorption of H on the hollow site on the surface layer in the presence of Cr, $E^{H(Cr)}_{ad}$, can be calculated using the following relation

$$E^{H(Cr)}_{ad} = (E_{Tot}(Fe_{N-2}Cr_2H_2) - E_{Tot}(Fe_{N-2}Cr_2) - E_{Tot}(H_2))/2, \qquad (2)$$

where $E_{Tot}(Fe_{N-2}Cr_2)$ and $E_{Tot}(Fe_{N-2}Cr_2H_2)$ are the total energies of the slabs containing (N-2) Fe atoms (here N=36) and two Cr atoms, respectively without and with H atoms on the surface site. The energies of solution of a Cr atom without ($E^{Cr}_{Sol}$) and with the presence of a H atom at the surface ($E^{Cr(H)}_{Sol}$) can be obtained from the following relations

$$E^{Cr}_{Sol} = (E_{Tot}(Fe_{N-2}Cr_2) - 2\mu_{Cr} + 2\mu_{Fe} - E_{Tot}(Fe_N))/2. \qquad (3)$$

$$E^{Cr(H)}_{Sol} = (E_{Tot}(Fe_{N-2}Cr_2H_2) - 2\mu_{Cr} + 2\mu_{Fe} - E_{Tot}(Fe_NH_2))/2. \qquad (4)$$

where $E_{Tot}(Fe_{N-1}CrH_2)$ is the total energy of a slab containing (N-1) Fe atoms, one Cr atom at the central layer, and two H atoms on the surface symmetrically placed in surface layers S, $E_{Tot}(Fe_NH_2)$ is the total energy of the slab without any Cr atoms but H atoms on the surfaces. $\mu_{Fe}$ and $\mu_{Cr}$ are chemical potentials of Fe and Cr respectively in their magnetic states.

The difference of the energy of solution with respect to its value in the bulk yields the energy of segregation of Cr, $E^{Cr(H)}_{seg}$, in the presence of H on the surface of Fe. This energy clearly does not depend upon the reference energies of Fe and Cr metals, and can be obtained directly from the following relation,

$$E^{Cr(H)}_{seg} = (E_{Tot}(Fe_{N-2}Cr_2H_2) - 2E_{Tot}(Fe_{N-1}CrH_2) + E_{Tot}(Fe_NH_2))/2. \qquad (5)$$

In our slab geometry, there are four hollow sites, located at (0.5 0.5) *a*, (0.5 1.5) *a*, (1.5 0.5) *a*, and (1.5 1.5) *a* with respect to the Cr atom placed at the origin (0.0 0.0) at the surface. Due to symmetry considerations, they are all identical. The energy of adsorption of H in the hollow site in the presence of Cr, with both being on the topmost surface layer is found to be somewhat more exothermic, -0.447eV/H atom than at the surface of pure Fe, -0.405eV/H atom. Similarly, the energy of solution of Cr at this layer is also slightly more exothermic, -0.127eV/Cr, compared to a value, -0.085eV/Cr in the absence of hydrogen on the surface. Thus both the energies of hydrogen adsorption and the energies of Cr solution are more exothermic when both are present together. This shows the synergistic effect of the interaction of these two elements. An interesting observation from these results is that the energy of adsorption of H is more exothermic by -0.042eV/H atom, and this value is exactly the same as the energy of solution of Cr that is also more exothermic by -0.042eV/Cr atom. In fact, this result is fully expected since the difference of the energies of solution of Cr with and without the presence of H (Eq. (4) and Eq.(3)) is exactly the same as the difference of the energies of H adsorption with and without the presence of Cr at the surface layer (Eq. (2) and Eq.(1)). We note that the H atom is adsorbed at a slightly higher vertical distance, 0.39Å, compared to the Cr free case. Also it is laterally displaced by ~0.09Å away from its initial position in the surface plane. In fact, the Cr atom is also displaced both vertically upwards by 0.05Å and laterally by -0.06Å from its initial position. Thus the H atom is at a distance longer by 0.15Å from the Cr atom in the surface plane, but at a slightly reduced distance, 0.34Å, in the vertical direction from the Cr atom. But on the whole, the H atom is at a considerably larger distance from the Cr atom, 2.18Å, than from a Fe atom in the surface plane, 2.06Å. As discussed above in the case of pure Fe, the adsorption energy of H is determined mainly from the interaction of the H atom with a single Fe atom in the subsurface layer just below it. We find that this distance is only marginally reduced with respect to its value in pure Fe. Thus the slight reduction in this bond length, that reinforces this bonding, coupled with the gain in the relaxation energy associated with the surface atoms, is responsible for the slightly more exothermic character of H adsorption and Cr solution in Fe when both are present on the topmost surface layer. In Table II, we have shown the values of the interlayer relaxations and the magnetic moments at the Fe sites at different surface layers, which are not found to be significantly affected by the presence of H atom at the surface. Similarly, the value of the magnetic moment at the Cr site, $-3.10\mu_B$, remains unaffected by the presence of H on the surface. The H coverage on the surface does not seem to change much the energy of hydrogen adsorption since with the full one monolayer coverage, meaning all four hollow sites occupied by the H atoms, we obtain an energy of H adsorption, -0.461eV/H atom, only slightly different from the value, -0.447eV, obtained above for a much lower coverage. But it does reveal once again the attractive nature of the H-H interaction on the surface of Fe in the presence of Cr, as was found above also for the case of pure Fe.

The situation is somewhat more complicated when H is on the hollow site on the topmost surface layer, and Cr is in the subsurface layer. As shown in Fig. 2, in this case there are three distinct hollow sites for H adsorption on the surface; one (denoted H0) directly above the Cr atom at an ideal distance of 0.5$a$, second (denoted H1) laterally displaced by (1.0 1.0) $a$ from this site, and the third (denoted H2), laterally displaced by (1.0 0.0) $a$ or (0.0 1.0) $a$ from the H0 site, where $a$ is the lattice constant of Fe. As shown in Table III, the adsorption energies in all three cases are found to be more exothermic than without Cr in the subsurface layer but the last configuration (H2) is found to be the most stable one, and the one immediately on the top of the Cr atom (H0) is the least preferred. The H0 site is the closest, and the H1 site the furthest away from the Cr atom situated in the subsurface layer while the H2 site lies in between. These results show that there is an optimal distance between the H atom at the surface and the Cr atom in the subsurface layer. The peculiarity of these sites arises from the complex nature of the interactions between the H atom on the surface and the Cr atom at the subsurface site, and the Fe atoms on the surface. It turns out that the ideal lattice site positions for the Cr atom in the subsurface layer are not the lowest energy sites in the presence of the H atom on the surface. In fact, the Cr atom in the subsurface layer is forced to displace itself laterally in its plane, by more than 0.03 Å, and this displaces the H atom and the surface Fe atoms also laterally (in the surface plane) by ~0.10Å and ~0.15Å respectively. The H-Cr distance also increases by ~0.04Å. These relaxations result in a lowering of the energy of hydrogen adsorption by approximately ~0.065eV/H in all three cases from the energy obtained without any lateral displacement of Cr, and a consequent lowering of the energy of Cr solution in the subsurface layer by this same amount, -0.065eV/Cr atom, as was discussed above on the basis of Eqs.(1)-(4). These energies have been given in Table III, and include all the lateral displacements induced by Cr lateral displacement due to hydrogen. Such lateral displacements are not found in the absence of hydrogen, and thus reflect the complex nature of the interaction of H with Cr in ferromagnetic Fe. For the most stable site, H2, we obtain a value, -0.526eV/H, for the energy of H adsorption in the presence of Cr, and this has to be compared with the value, -0.405eV/H, obtained without Cr, and corresponds to an energy lowering of -0.121eV/H, due to Cr in the subsurface layer and H on the surface layer. This lowers the energy of Cr incorporation in this layer by the same amount, from 0.192eV in the absence of H to 0.071eV with H on the topmost surface layer. In this configuration (H2), the H atom is closer to the surface, practically at the same distance as in the Cr free case, as compared to the other two configurations H0 and H1, as shown in Table III. The interlayer relaxations and the related magnetic moments are given in Table IV. There are only minor differences with respect to the H free case, and do not provide additional information concerning the energy lowering due to the presence of H on the surface, and Cr in the subsurface layer. The distance of the H atom from the nearest neighbor Fe atom in the subsurface layer hardly changes from its value in the Cr free case. Thus this additional energy lowering, compared to the Cr free case, clearly occurs due to the displacement of the Cr and H atoms and the associated displacements of the surface Fe atoms. A Cr atom in the subsurface layer has four nearest neighbor Fe atoms on the topmost surface layer, and another four in the sub-subsurface layer underneath it. It turns out that the four Fe atoms on the surface get split into two groups in the presence of H on the surface, two with a Cr-Fe distance of 2.41Å and another two with a longer distance of 2.54Å, while the other four Fe atoms in the sub-subsurface layer all remain at the same distance 2.40Å. Such a splitting was not found in the absence of H in Fe-Cr with Cr located in the subsurface layer, or in the absence of Cr in the subsurface layer with H at the surface of Fe.

These relaxations affect the nature of the interaction of the H atom with its nearest neighbor Fe atom just underneath it in the subsurface layer. In Fig. 3, we have shown the total (the sum of spin up and spin down) densities of states (DOS) at this Fe atom site, both with and without Cr. We see clearly that, in the presence of Cr, the low energy portion of the DOS is pulled toward lower energies, indicating stronger stability of both H and Cr in Fe when both are present together. In Table V, we have shown the solution and segregation energies of a Cr atom to the topmost surface layer and the subsurface layer. The solution energy of Cr in the bulk (central layer), -0.163eV, was taken from our previous work [14]. We see that the adsorption of the H atom on the topmost surface layer does facilitate the segregation of Cr to the surface since it lowers considerably the barrier for surface segregation. If one ignores the subsurface layer, the segregation energy to the topmost surface layer is nearly zero, implying segregation of Cr at very low temperatures. On the other hand, the subsurface layer still does present a barrier, but this is nevertheless much smaller, 0.235eV, compared to the value 0.355eV in the H free case.

## IV. CONCLUSION

In general, the solution energy, in solid solution forming dilute alloys, increases more or less smoothly from the bulk towards the surface layers, and the difference of the solution energy at the topmost surface layer relative to the bulk determines the segregation energy. The dilute Fe-Cr system is anomalous since there is a large sudden (endothermic) increase in the solution energy, and hence the segregation energy, in the subsurface layer just below the topmost surface layer after which it drops again to a very low value at the surface. Thus the topmost surface layer no longer controls the segregation behavior of Cr in dilute Fe-Cr alloys, which is now governed instead by the subsurface layer. This could result in an incorrect interpretation of the data since one usually determines the segregation energy from the difference of the solution energies in the topmost surface layer and in the bulk, and the subsurface layer is never considered. We have shown that this anomaly is considerably reduced, but not eliminated, by the presence of H adsorbed on the surface of Fe. This is due to the synergistic nature of coupling between H and Cr. The energy of adsorption of H on the surface of Fe tends to become more exothermic in the presence of Cr, and this in turn reduces considerably the endothermic nature of Cr solution in the subsurface layer. The synergistic interaction of H with Cr in Fe is very complex. The H atom at the surface of Fe displaces laterally the Cr atoms in the subsurface layer that in turn displaces the surface Fe atoms, resulting in a lowering of the solution energies of both H and Cr in Fe. Such displacements are not found in the absence of hydrogen. The Fe atoms on the surface layer play the key role in lowering the segregation barrier in the subsurface layer.

ACKNOWLEDGMENTS

We would like to thank IDRIS (Institut du Développement et des Ressources en Informatique Scientifique) for providing us access to the high power computing (HPC) resources of GENCI

$email : michele.gupta@u-psud.fr

TABLE I  Interlayer relaxations (Å) and magnetic moments ($\mu_B$) of Fe in pure Fe, and in the presence of H, in the surface (S), subsurface (S1), an sub-subsurface (S2) layers. The hydrogen atom is adsorbed at a vertical height of 0.38Å with an energy of -0.405eV at the fourfold coordinated hollow site

---

| Layer | Interlayer Relaxation | | Magnetic Moment ($\mu_B$) | |
|---|---|---|---|---|
| | $Fe_{36}$ | $Fe_{36}H_2$ | $Fe_{36}$ | $Fe_{36}H_2$ |
| S  | -0.021 | -0.024 | 2.83 | 2.75 |
| S1 | +0.044 | +0.042 | 2.32 | 2.21 |
| S2 | +0.013 | 0.000  | 2.38 | 2.36 |

---

TABLE II Interlayer relaxations (Å) and magnetic moments ($\mu_B$) of Fe in surface (S), subsurface (S1), and sub-subsurface layers, with Cr at the topmost surface layer, without and with H at the fourfold coordinated hollow site. The H atom is adsorbed at a vertical height of 0.39Å with an energy of -0.447eV. The Cr atom carries a magnetic moment of -3.10$\mu_B$, both without and with H at the surface. The energy of solution of Cr at this layer is found to be -0.127eV.

---

| Layer | Interlayer Relaxation | | Magnetic Moment ($\mu_B$) | |
|---|---|---|---|---|
| | $Fe_{34}Cr_2$ | $Fe_{34}Cr_2H_2$ | $Fe_{34}Cr_2$ | $Fe_{34}Cr_2H_2$ |
| S  | -0.066 | -0.062 | 2.71 | 2.64 |
| S1 | +0.045 | +0.044 | 2.15 | 2.11 |
| S2 | 0.000  | 0.000  | 2.39 | 2.38 |

---

TABLE III. Energies of solution of a H atom at different hollow sites at the (1 0 0) surface of Fe without Cr, $E_{sol}^{H}$, and in the presence of Cr at the subsurface layer, $E_{sol}^{H(Cr)}$, and the energies of solution of Cr in the subsurface layer without H, $E_{sol}^{Cr}$, and with H in the topmost surface layer, $E_{sol}^{Cr(H)}$. The distances of H from the topmost surface layer D(H-S) and the magnetic moments (MM) of Cr are listed.

|  | H0 | H1 | H2 |
|---|---|---|---|
| $E_{sol}^{H}$ (eV) | -0.405 | -0.405 | -0.405 |
| $E_{sol}^{Cr}$ (eV) | 0.192 | 0.192 | 0.192 |
| $E_{sol}^{H(Cr)}$ (eV) | -0.426 | -0.467 | -0.525 |
| $E_{sol}^{Cr(H)}$ (eV) | 0.171 | 0.130 | 0.072 |
| D(H-S) (Å) | 0.43 | 0.40 | 0.38 |
| MM (Cr) ($\mu_B$) | -1.53 | -1.57 | -1.84 |

TABLE IV Interlayer relaxations (Å) and magnetic moments ($\mu_B$) of Fe in surface (S), subsurface (S1), and sub-subsurface (S2) layers with Cr at the subsurface layer, without and with H at the most stable fourfold coordinated hollow site H2. The H atom is adsorbed at a vertical height of 0.38Å with an energy of -0.526eV. The Cr atom carries a magnetic moment of -1.80 $\mu_B$ without and -1.84 $\mu_B$ with H at the surface.

---

| Layer | Interlayer Relaxation | | Magnetic Moment ($\mu_B$) | |
|---|---|---|---|---|
|  | $Fe_{34}Cr_2$ | $Fe_{34}Cr_2H_2$ | $Fe_{34}Cr_2$ | $Fe_{34}Cr_2H_2$ |
| S | -0.038 | -0.024 | 2.66 | 2.56 |
| S1 | +0.055 | 0.045 | 2.24 | 2.21 |
| S2 | 0.011 | 0.012 | 2.28 | 2.26 |

---

TABLE V Energies of solution, $E_{Sol}$, and segregation, $E_{Seg}$, of Cr in surface (S), subsurface (S1), and central (C) layers without and with H at the topmost surface layer. The value of the magnetic moment, MM, at the Cr site (aligned opposite to Fe atoms) is also given.

| Layer | $E_{Sol}$ (eV) without H | $E_{Sol}$ (eV) with H | $E_{Seg}$ (eV) without H | $E_{Seg}$ (eV) with H | MM Cr ($\mu_B$) without H | MM Cr ($\mu_B$) with H |
|---|---|---|---|---|---|---|
| S  | -0.085 | -0.127 | 0.078 | 0.036 | -3.10 | -3.10 |
| S1 | 0.192  | 0.072  | 0.355 | 0.235 | -1.80 | -1.84 |
| C  | -0.163 | -0.163 | 0.000 | 0.000 | -1.60 | -1.60 |

Figure Captions:

FIG. 1. (Color online) (a) perspective view of octahedral sites (shown in blue and green) in bulk Fe. The bcc lattice of Fe is shown by black circles. The octahedral sites denoted O1 (shown by blue circles) and O2 (shown by green circles) are identical in pure bulk Fe, (b) perspective view of tetrahedral sites (shown in yellow) in bulk Fe, (c) top view of the (1 0 0) surface of Fe. The octahedral sites of the bulk become different on the surface, with O1 becoming the hollow (H) site and O2 the bridge (B) site. The tetrahedral sites are shown by small yellow circles.

FIG. 2. (Color online) With Cr (shown in red) situated in the subsurface layer, the hollow (H) sites for H adsorption on the (1 0 0) surface of Fe, shown by crosses, split into three groups H0 (immediately above the Cr atom), H1 (displaced laterally by (1.0 1.0) $a$ from the H0 site, and H2 (displaced laterally from the H0 site by (1.0 0.0) $a$, or equivalently by (0.0 1.0) $a$.

FIG. 3. Total (sum of spin up and spin down) densities of states (DOS) at the subsurface Fe site just underneath the H atom on the surface, in pure Fe (top panel), and with Cr in the subsurface layer (bottom panel). The Fermi level is located at zero of the energy scale.

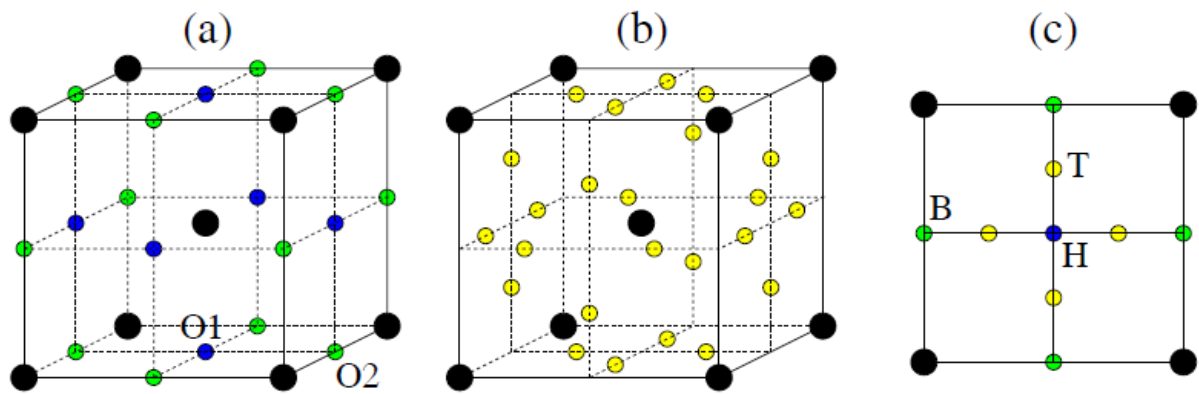

FIG. 1. (Color online) (a) perspective view of octahedral sites (shown in blue and green) in bulk Fe. The bcc lattice of Fe is shown by black circles. The octahedral sites denoted O1 (shown by blue circles) and O2 (shown by green circles) are identical in pure bulk Fe, (b) perspective view of tetrahedral sites (shown in yellow) in bulk Fe, (c) top view of the (1 0 0) surface of Fe. The octahedral sites of the bulk become different on the surface, with O1 becoming the hollow (H) site and O2 the bridge (B) site. The tetrahedral sites are shown by small yellow circles.

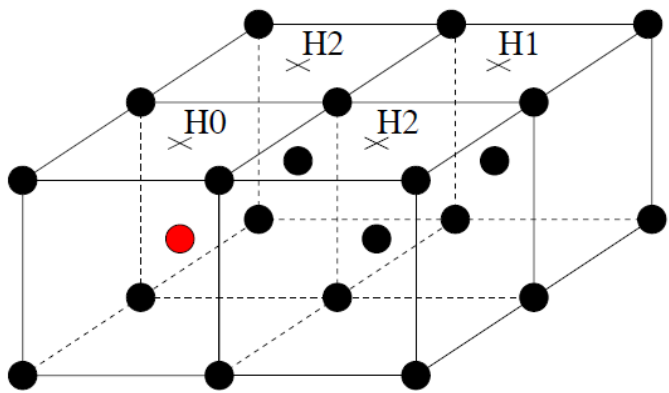

FIG. 2. (Color online) With Cr (shown in red) situated in the subsurface layer, the hollow (H) sites for H adsorption on the (1 0 0) surface of Fe, shown by crosses, split into three groups H0 (immediately above the Cr atom), H1 (displaced laterally by (1.0 1.0) $a$ from the H0 site, and H2 (displaced laterally from the H0 site by (1.0 0.0) $a$, or equivalently by (0.0 1.0) $a$.

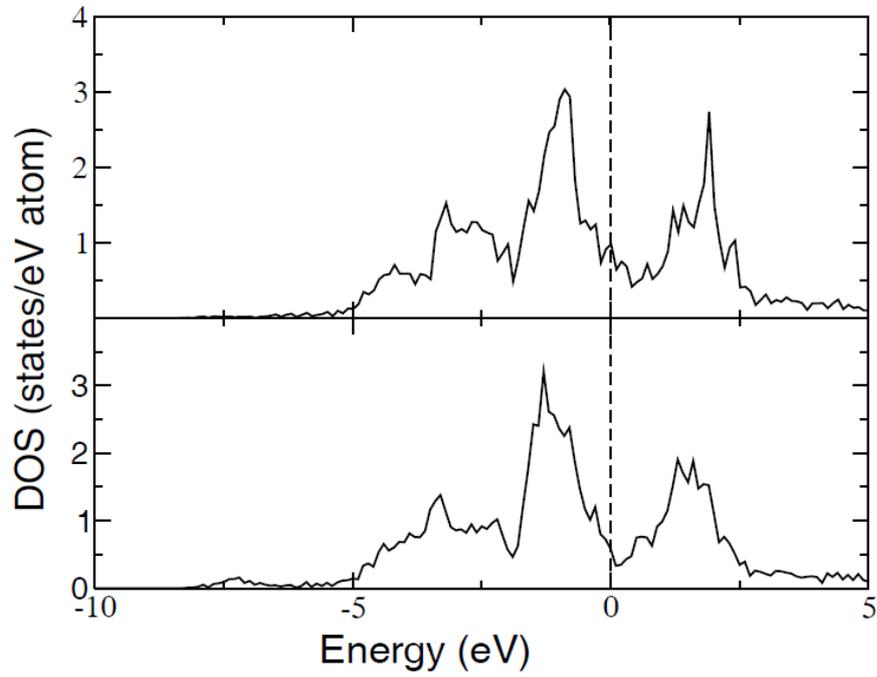

FIG. 3. Total (sum of spin up and spin down) densities of states (DOS) at the subsurface Fe site just underneath the H atom on the surface, in pure Fe (top panel), and with Cr in the subsurface layer (bottom panel). The Fermi level is located at zero of the energy scale.